# A power-efficient integrated lithium niobate electro-optic comb generator

Ke Zhang[1,†], Wenzhao Sun[1,2,3,†], Yikun Chen[1], Hanke Feng[1], Yiwen Zhang[1], Zhaoxi Chen[1], Cheng Wang[1,*]

[1] Department of Electrical Engineering & State Key Laboratory of Terahertz and Millimeter Waves, City University of Hong Kong, Kowloon, Hong Kong, China
[2] Centre of Internet of Things, City University of Hong Kong Dongguan Research Institute, Dongguan, China
[3] Centre of Information and Communication Technology, City University of Hong Kong Shenzhen Research Institute, Shenzhen, China

[†] These authors contributed equally
[*] Corresponding author: cwang257@cityu.edu.hk

## Abstract

Integrated electro-optic (EO) frequency combs are essential components for future applications in optical communications, light detection and ranging, optical computation, sensing and spectroscopy. To date, broadband on-chip EO combs are typically generated in high-quality-factor micro-resonators, while the more straightforward and flexible non-resonant method, usually using single or cascaded EO phase modulators, often requires high driving power to realize a reasonably strong modulation index. Here, we show that the phase modulation efficiency of an integrated lithium niobate modulator could be dramatically enhanced by passing optical signals through the modulation electrodes for a total of 4 round trips, via multiple low-loss $TE_0/TE_1$ mode multiplexers and waveguide crossings, reducing electrical power consumption by more than one order of magnitude. Using devices fabricated from a wafer-scale stepper lithography process, we demonstrate a broadband optical frequency comb featuring 47 comb lines at a 25-GHz repetition rate, using a moderate RF driving power of 28 dBm (0.63 W). Leveraging the excellent tunability in repetition rate and operation wavelength, our power-efficient EO comb generator could serve as a compact low-cost solution for future high-speed data transmission, sensing and spectroscopy, as well as classical and quantum optical computation systems.

## Introduction

Optical frequency combs (OFCs), featuring discrete, equally spaced optical frequency components[1], are excellent building blocks for optical communication[2,3], light detection and ranging (LiDAR)[4], optical computation[5], optical clocks[6,7] and spectroscopy[8,9]. Chip-scale OFCs, taking advantage of recent developments in photonic integrated circuits (PICs), could further allow the above functions achieved in a compact and cost-effective manner. To date, most on-chip frequency comb generators are based on semiconductor mode-locked lasers or nonlinear Kerr effect ($\chi^{(3)}$)[10-15]. The former approach employs passive or active mode locking schemes to directly achieve mode-locked lasing states in integrated III–V gain platforms, leading to smaller footprint and improved wall-plug efficiencies compared to traditional solid-state or fiber-based mode-locked lasers[10-13]. The latter process originates from phase-locked cascaded four-wave-mixing (FWM) in ultrahigh-quality microresonators[14,15], resulting in low-noise OFCs with broad spectral span achieved in various photonic platforms, such as silicon (Si)[16], silicon dioxide ($SiO_2$)[17], silicon nitride (SiN)[18], lithium niobate (LN)[19], silicon carbide (SiC)[20], aluminum nitride (AlN)[21] and so on. Historically, there also exists a well-known and widely adopted alternative approach, namely electro-optic (EO) comb generation, to produce combs with high optical powers and widely tunable repetition rates via cascaded sideband generation processes in one or multiple EO phase modulators. Most EO combs to date, however, still rely on discrete LN modulators that are bulky and rather inefficient, leading to systems that are on table-top scales and consume substantial RF powers[22,23].

The recent technological advances in the LN-on-insulator (LNOI) photonic platform have inspired a renaissance for EO combs to be achieved in chip-scale systems that are more efficient, more compact and lower cost[24-30]. In the LNOI platform, a sub-micron-thick LN thin film is bonded on top of $SiO_2$ dielectric substrate, resulting in much better light confinement, and substantially improved EO modulation efficiencies[24-36]. As a result, breakthroughs have been made on LNOI-based on-chip EO comb generators in aspects of spectral breadth, light conversion efficiency and comb line flatness[24-30]. For example, EO combs generated using a single LN resonator can cover the entire telecom L band with over 900 lines[24], and light conversion efficiency can be further improved to 30% using a two-resonator system[27]. Leveraging the high quality-factor (Q-factor), strong microwave-optic field overlap, and engineered dispersion, optical signals can pass through the EO modulation area many times, leading to a strong and cascaded sideband generation process, and in turn a broadband comb span. However, these high-Q-resonator-based EO combs usually suffer from limited tuning ranges in both the RF repetition rate and the optical operation wavelength due to the narrow resonance linewidths, hurdling their practical applications, e.g. in frequency-modulated LiDAR systems[37,38]. On the contrary, the more conventional and straightforward non-resonant EO comb generation scheme offers much more flexibility in selecting and tuning the repetition rate and operation wavelength on demand. Cascading an amplitude and a phase modulator could further allow the generation of flat-top EO combs[29,30]. Unfortunately, due to the non-resonant structure, light passes through the EO modulation area only once for each modulator, leading to a relatively weak EO modulation effect and high driving voltage (power consumption) needed. Although longer metal electrodes can be utilized to induce a larger phase shift at low frequencies, the increased microwave loss and subsequently lowered EO bandwidth ultimately limit the maximally achievable modulation index at high frequencies (> 20 GHz). For instance, EO combs featuring 40 comb lines can be generated using a single LNOI phase modulator, which however requires a high RF power of 3.1 W[28]. More recently, it has been demonstrated that, by recycling optical signals through both modulation areas of a typical ground-signal-ground (GSG) electrode structure, a double-pass EO phase modulator could provide a doubled phase modulation efficiency, leading to the generation of 67 comb lines spaced at 30 GHz using an RF power of 4.0 W[30].

In this article, we demonstrate a power-efficient LNOI-based EO comb generator, where the phase modulation efficiency is increased by a factor of 4 using a multi-loop design, leading to a subsequent reduction of electrical power consumption by 16 times. This power-efficient characteristic benefits from passing optical signals through the modulation electrodes for a total of 4 round trips with the assistance of multiple low-loss $TE_0/TE_1$ mode conversion processes. We show the generation of broadband optical frequency combs with 47 comb lines at a 25-GHz repetition rate using a relatively low RF driving power of 28 dBm, while maintaining excellent tunability in both the repetition rate and operation wavelength.

## Results

**Overall design** Fig. 1(a) shows the schematic of our EO comb generator, where CW pump light (fundamental transverse-electric mode, $TE_0$) from the left hand side travels through the phase modulation region (yellow electrode area) for the first EO modulation, then loops back and converts into a $TE_1$ mode via an adiabatic $TE_0/TE_1$ mode multiplexer (red dashed arrow in Fig. 1(b)). This $TE_1$ mode light, which is orthogonal to the $TE_0$ mode in the first pass, subsequently goes through the phase modulation region for a second time, and is converted back to $TE_0$ mode in the lower branch of a second $TE_0/TE_1$ mode multiplexer [red dashed arrow in Fig .1(c)]. The adiabatic mode multiplexer is designed such that $TE_0$ mode in the main branch stays in $TE_0$ mode throughout the mode evolution process (details to be discussed next). The twice modulated optical signal is further guided back to the bottom modulation region of the GSG transmission line. Similar to the top part, light passes through the modulation area for a third and a fourth time, in $TE_0$ and $TE_1$ mode respectively, ultimately reducing the drive voltage (power) of this comb generator by a factor of 4 (16). Importantly, the microwave-optic phase-matching condition needs to be satisfied in two aspects in our circulating waveguide design. First, similar to a normal EO modulator, the microwave phase velocity should be matched with the optical group velocity throughout the modulation region, which is achieved using a GSG traveling-wave electrode design and an engineered material stack [Fig. 1(d)], similar to those used in previous literature[31,39]. The electrode length in our current device is 1 cm to avoid excessive RF losses at high frequencies (measured EO 3-dB bandwidth > 40 GHz). Second, the optical signals after each round trip should be precisely delayed, as shown in the microscope image of the full device [Fig. 1(e)], such that they experience the same RF phase in each modulation step for our target RF frequency of 25 GHz. Finally, the 4-time modulated signal outputs toward the right hand side through a carefully designed waveguide crossing [Fig. 1(f)] that features low losses of -0.18 dB in the *y*-crystal direction and -0.10 dB in the *z*-crystal direction (see **Supplementary S1**). Figure 1(b,c,f,g) shows the scanning electron microscope (SEM) images of the adiabatic mode multiplexers, waveguide crossing and the EO modulation region. The x-cut LNOI devices are fabricated using a wafer-scale UV stepper lithography-based process with high uniformity and repeatability [Fig. 1(h)], more details of which could be found in "**Methods,** Wafer-scale LNOI device fabrication".

**Adiabatic mode multiplexer** A key enabling component for our multi-loop EO comb generator is an efficient, low-crosstalk and low-loss $TE_0/TE_1$ mode multiplexer. This is achieved using a broadband and robust adiabatic coupler design as shown in Fig. 2(a), where incoming $TE_0$ and $TE_1$ modes from the left hand side are demultiplexed into the upper output branch (branch 1) and the lower output branch (branch 2), respectively, both in $TE_0$ mode. Insets of Fig. 2(a) show the simulated eigenmode profiles ($|E_z|$) at different positions of the adiabatic mode multiplexer. This mode-multiplexing function is achieved based on a 550-µm-long adiabatic coupling region, where branch 1 gradually narrows while branch 2 widens along the optical propagation direction, leading to a drop of the effective index ($n_{eff}$) of $TE_1$ mode in branch 1 and an increase for $TE_0$ mode in branch 2 [Fig 2(b)]. When the two modes cross over each other in $n_{eff}$, a substantial avoided crossing ($\Delta n_{eff}$ = 0.01) is generated between the super-modes of the two-waveguide-coupled system, since the two optical modes share the same polarization and a finite field overlap over the coupling gap (500 nm) in our device. As a result, the optical energy of $TE_1$ mode in branch 1 is gradually transferred to and in the end totally converted to $TE_0$ mode in branch 2, following the red curve in Fig. 2(b). Meanwhile, $TE_0$ mode in branch 1 stays the lowest-order mode of the coupled-waveguide system throughout the coupling region, and remains in $TE_0$ mode in branch 1 at the output end, shown as the green curve in Fig. 2(b). To evaluate the performance of our adiabatic mode multiplexer, we design and fabricate a 2 × 2 cascaded structure using the same fabrication method as the actual comb generator, as shown in the inset of Fig. 2(c). Ideally, incoming light from port 1 (in $TE_0$ mode) will stay decoupled and output from port 3, whereas input light from port 2 will first be converted to $TE_1$ mode in the mode multiplexed middle section, and finally output from port 4. Figure 2(c) shows the measured *S*-parameters of this test structure, where $S_{ij}$ refers to the ratio between the output power of port *i* and the input power of port *j*. The device shows low optical losses of < 0.4 dB ($S_{31}$ and $S_{42}$) and high extinction ratios of > 20 dB over a broad wavelength range from 1530 nm to 1630nm, which is important to the wide operation wavelength window of our EO comb generator. Moreover, our simulation results (see **Supplementary S2**) suggest that the adiabatic multiplexer design is robust against variations in waveguide coupling gap, cladding material, waveguide width, and sidewall angle[33,35].

**Broadband EO comb generation** We demonstrate broadband EO comb generation at low RF drive power using our multi-loop on-chip EO comb generator. The corresponding experimental setup is shown in Fig. 3(a) (see "**Methods**, EO comb generation" for more details). For better comparison and visualization of the enhanced comb generation process in our device, we fabricate and test four types of phase modulators [top row of Fig. 3(b)]: (I) a single-pass phase modulator; (II) a double-pass phase modulator making use of both top and bottom modulation areas, both in fundamental $TE_0$ mode; (III) another type of double-pass phase modulator, with $TE_0/TE_1$ mode multiplexing but only going through the top modulation region; (IV) our quadra-pass phase modulator. Figure 3(b) shows the simulated (middle row) and measured (bottom row) EO comb spectra using the above four types of comb generators driven at the same RF power of 28 dBm (630 mW) at 24.95 GHz, which corresponds to a peak voltage of $V_p$ = 7.94 V and a modulation index of ~ 1.05π for a simple phase modulator (DC half-wave voltage $V_\pi$ = 6.20 V, RF $V_\pi$ = 7.56 V at 25 GHz). A total of 15 comb lines are generated in the single-pass device, in line with the numerically simulated spectrum [Fig. 3(b), Type I]. The EO comb span is substantially broadened in the double-pass cases, to 25 lines in the type-II device and 27 lines in the type-III device, with increased effective modulation index of ~ 2.1π. Finally, our quadra-pass EO comb generator (type-IV), a combination of type-II and type-III, shows a total of 47 measured comb lines [Fig. 3(b), Type IV], corresponding to a 4× enhancement of the modulation index (~ 4.2π) and a subsequent reduction of electrical power consumption by 16 times. We also measure the generated comb spectra of our device at different RF driving power levels [Fig. 3(c)], showing 7, 16 and 47 comb lines at RF powers of 14 dBm, 23 dBm and 28 dBm, respectively. Our simulation results indicate that an even broader EO comb [purple spectrum in Fig. 3(d)] with 87 comb lines could be generated using our device if driven at 36 dBm (~ 10.5π equivalently), a similar power as that used in previous literature[30] for a 30-GHz EO comb, spanning an optical bandwidth of close to 20 nm. Importantly, the dramatically reduced RF power consumption in our multi-pass EO comb generator does not come at the cost of substantially increased optical loss, as the measured optical transmission spectra of the four device types in Fig. 3(d) show. Although waveguides with a total of ~12 cm long and structures such as waveguide crossing and mode multiplexers (8 passes in total) are used, our EO comb generator (blue curve) only induces a total on-chip optical loss of ~ 4 dB, benefiting from the low-loss and high-uniformity wafer-scale LN photonic platform used (see "**Methods**, Optical loss characterization" for more details). Further improving the waveguide propagation loss to < 0.1 dB/cm for both $TE_0$ and $TE_1$ modes[33] could allow even lower total on-chip optical loss of < 1 dB.

**RF and optical tuning range** Thanks to the non-resonant configuration of our design, our power-efficient EO comb generator could efficiently operate within a broad range of RF and optical frequencies. To further characterize the frequency response of our EO comb generator at different RF frequencies, we specially design and fabricate a Mach–Zehnder interferometer (MZI) [Fig. 4(a)] formed by two identical multi-loop phase modulators on the two arms (see **Supplementary S3**). Electrical signals added to one of the two phase modulators therefore are translated into amplitude modulation, allowing us to directly measure the EO response of our multi-loop phase modulator [red dots in. Fig. 4(b), see "**Methods,** EO response and RF $V_\pi$ characterizations" for more details]. The directly measured EO $S_{21}$ response of our device shows excellent agreement with our simulation prediction (grey curve) within the measured frequency range (up to 10 GHz, limited by our photodetector, PD), indicating that the desired optical delays have been precisely achieved in our device. The device EO $S_{21}$ response shows oscillation behaviors as a function of frequency, reaching the maximal phase modulation efficiency every 4.6 GHz, since the second RF-optic phase matching condition is only satisfied at certain RF frequencies for a fixed loop delay. At higher frequencies, the EO responses are measured by driving the multi-loop phase modulator at the small-signal limit and extracting the modulation sideband-to-carrier ratios at different frequencies using an optical spectrum analyzer (OSA, see "**Methods,** EO response and RF $V_\pi$ characterizations" for more details), which shows larger measurement uncertainties but also agrees reasonably well with the theoretically predicted trend. The maximally achievable EO $S_{21}$ values are bounded by a slowly rolling-off envelope (grey dash line) as a result of the intrinsic bandwidth limit of the phase modulator, which features a 3-dB bandwidth larger than 40 GHz in the current device. We further extract and plot the measured RF $V_\pi$ values of our device at various RF frequencies in Fig. 4(c), showing a minimal RF $V_\pi$ value of 1.90 V at 24.95 GHz and a 3-dB tuning range of 1.4 GHz, from 24 GHz to 25.4 GHz [defined as the range within which RF $V_\pi$ is no more than $\sqrt{2}V_{\pi,min}$, Fig. 4(d)]. The large RF frequency tuning range as compared with those of resonator-based EO combs[24] are important for practical applications that requires tunable repetition rates and/or are sensitive to environmental drifts. Apart from operating in the vicinity of 25 GHz, our EO comb generator can also efficiently function at other periodically occurring maximal-EO-response frequencies, e.g. near 20 GHz and 30 GHz [Fig. 4(e)]. At all three phase-matched frequencies (i.e. 20GHz, 25GHz, 30GHz), our device is able to generate more than 45 comb lines, showing great flexibility in input RF frequency. Finally, we show that our power-efficient EO comb generator is wavelength insensitive by pumping the device at different laser wavelengths, from 1520 nm to 1610 nm. As shown in Fig. 4(f), all measured comb spectra feature more than 45 lines, thanks to the broadband nature of our phase modulator, mode multiplexers and waveguide crossing, indicating a high degree of freedom in tuning the center wavelength of the generated EO combs.

## Discussions

Our power-efficient broadband integrated EO comb generator is enabled by a combination of the multi-pass phase modulator design that increases the modulation efficiency by 4 folds, low-loss and low-crosstalk $TE_0/TE_1$ mode multiplexers and waveguide crossing that allow for coherent phase accumulation between loops, as well as a reliable wafer-scale device fabrication platform that yields photonic devices and circuits with high uniformity, repeatability and low loss. The low RF $V_\pi$ of 1.90 V at 24.95 GHz in our current device could potentially be further reduced using even longer electrodes[30]. The 25-GHz-spaced, wavelength tunable EO comb could readily be matched with standard dense wavelength-division multiplexing (DWDM) grids (e.g. 50 GHz) in optical communications by filtering out half of the comb lines using, e.g. a microring resonator. Larger comb spacing could also be realized by directly driving the comb generators at higher RF frequencies, potentially using a capacitive-loaded electrode design with substantially lower RF losses[40]. On the other hand, the highly scalable stepper-lithography fabrication process could allow for larger-scale comb-based PICs with advanced system functionalities. For example, the multi-pass phase modulator could be further integrated with an amplitude modulator to achieve flat-top broadband EO combs at low RF power, important for practical DWDM systems. Dispersion elements can also be added after our EO comb generator for ultrashort pulse generation[30]. Our power-efficient EO comb generator could become a key building block in future chip-scale frequency comb systems for high-speed optical communications, LiDAR, sensing and spectroscopy applications.

Note added: During the preparation of this manuscript, we note a pre-print manuscript[41] posted by another group that makes use of a similar mode-multiplexing approach but only passes through one of the two modulation areas

## Methods

**Wafer-scale LNOI device fabrication** Our devices are fabricated from a commercially available x-cut LNOI wafer (NANOLN), which consists of a 500-nm LN thin film, a 2-μm buried $SiO_2$ layer, and a 500-μm silicon substrate. $SiO_2$ is first deposited on the surface of 4-inch LNOI wafter using plasma-enhanced chemical vapor deposition (PECVD). Micro/nano-structures are then directly patterned on the entire wafer using a ASML UV Stepper lithography system (NFF, HKUST) die by die (1.5 cm x 1.5 cm) with a resolution of 500 nm. Next, a reactive ion etching (RIE) system is used to transfer patterns from photoresist layer to $SiO_2$ layer and subsequently to LN wafer, leading to a 250 nm rib waveguide and a 250 nm LN slab. 1-cm-long metal electrodes are fabricated using a sequence of photolithography, thermal evaporation, and lift-off process, resulting in a 520-nm-thick metal layer with an electrode gap of 5.5 μm and a signal electrode width of 43 μm. $SiO_2$ is deposited by PECVD as cladding layer of the device. Finally, chips are cleaved and the facets are carefully polished for end-fire coupling.

**EO comb generation** CW pump light from a tunable telecom laser (Santec TSL-550) first goes through a fiber polarization controller (FPC) to ensure TE mode excitation. Light is coupled into and out from our chip using lensed fibers. RF signals (Anritsu, MG3697C) are amplified using a medium power GaAs amplifier (Pasternack, PE15A4021), before being delivered to the input port of the traveling-wave electrodes through a high-speed GSG probe. The output port of the electrodes is terminated with a 50-Ω load [Fig. 1(a)]. Finally, the generated EO comb spectra are collected using an optical spectrum analyzer (OSA).

**Optical loss characterization** Optical loss here is non-trivial due to our long and complex optical path design. The full device consists of 12-cm-long waveguides (10 cm in $TE_0$ mode and 2 cm in $TE_1$ mode), one waveguide crossing, and four mode multiplexers (8 passes in total), all of which contribute to the total on-chip optical loss. Based on the measured Q-factors of micro-resonators on the same chip, the waveguide loss is estimated to be ~ 0.2 dB $cm^{-1}$ for $TE_0$ mode and 0.4 dB $cm^{-1}$ for $TE_1$ mode, leading to a total waveguide loss of ~ 2.8 dB at 1550-nm wavelength. The waveguide crossing loss is evaluated by cut-back measurements of devices with different numbers of crossings (**Supplementary S2**), showing an average loss of 0.18 dB in y-crystal direction and 0.10 dB in z-crystal direction, leading to a total of 0.28 dB optical loss in the full device. The mode multiplexers exhibit high conversion efficiencies and low crosstalk [Fig. 2(c)], contributing to a total loss < 1 dB. Therefore, the total on-chip optical loss is estimated to be ~ 4 dB near 1550 nm, which matches with our measured transmission spectrum [Fig. 3(d)]. Except the on-chip optical loss, we note that the end-fire coupling system also induces optical loss of ~ 5 dB per facet.

**EO response and RF $V_\pi$ characterizations** A vector network analyzer (VNA) is used to measure the EO response of the MZI device, as shown in Fig. 4(a). The MZI is biased at the quadrature point while RF signals from the VNA are delivered to one of the multi-loop phase modulators. The majority (90%) of output optical signal is then sent to a high-speed photodetector (Newport), which translates the signal back to electrical signals for EO $S_{21}$ response measurement using the VNA. The remaining 10% is delivered to a 125-MHz photodetector to monitor the intensity of output light. At higher RF frequencies, the RF $V_\pi$ of our EO comb generator is characterized by monitoring the power ratio between optical pump and sideband signals using an OSA. The device is driven by a small RF signal (< 10 dBm) such that the system operates in the linear small-signal regime. RF frequency is swept from 18 GHz to 40 GHz, with a step of 200 MHz, to obtain the RF $V_\pi$ values at various frequencies.

## Acknowledgements

This work is supported in part by the National Natural Science Foundation of China (61922092); Research Grants Council, University Grants Committee (CityU 11212721, N_CityU113/20); Croucher Foundation (9509005); City University of Hong Kong (9610402, 9610455). We acknowledge Nanosystem Fabrication Facility (CWB) of HKUST for the device/system fabrication.

## Author contributions

K.Z. and C.W. conceived the idea in collaboration with the other co-authors. K.Z. designed the full device layout. K.Z., W.S., H.F. and Z.C. fabricated the device. K.Z. and Y.C. performed numerical simulations. K.Z., Y.C. and Y.Z. carried out the device characterization. K.Z. wrote the manuscript in discussion with all authors. C.W. supervised the project. All authors reviewed and approved the final manuscript.

## Competing interests

The authors declare no competing interests.

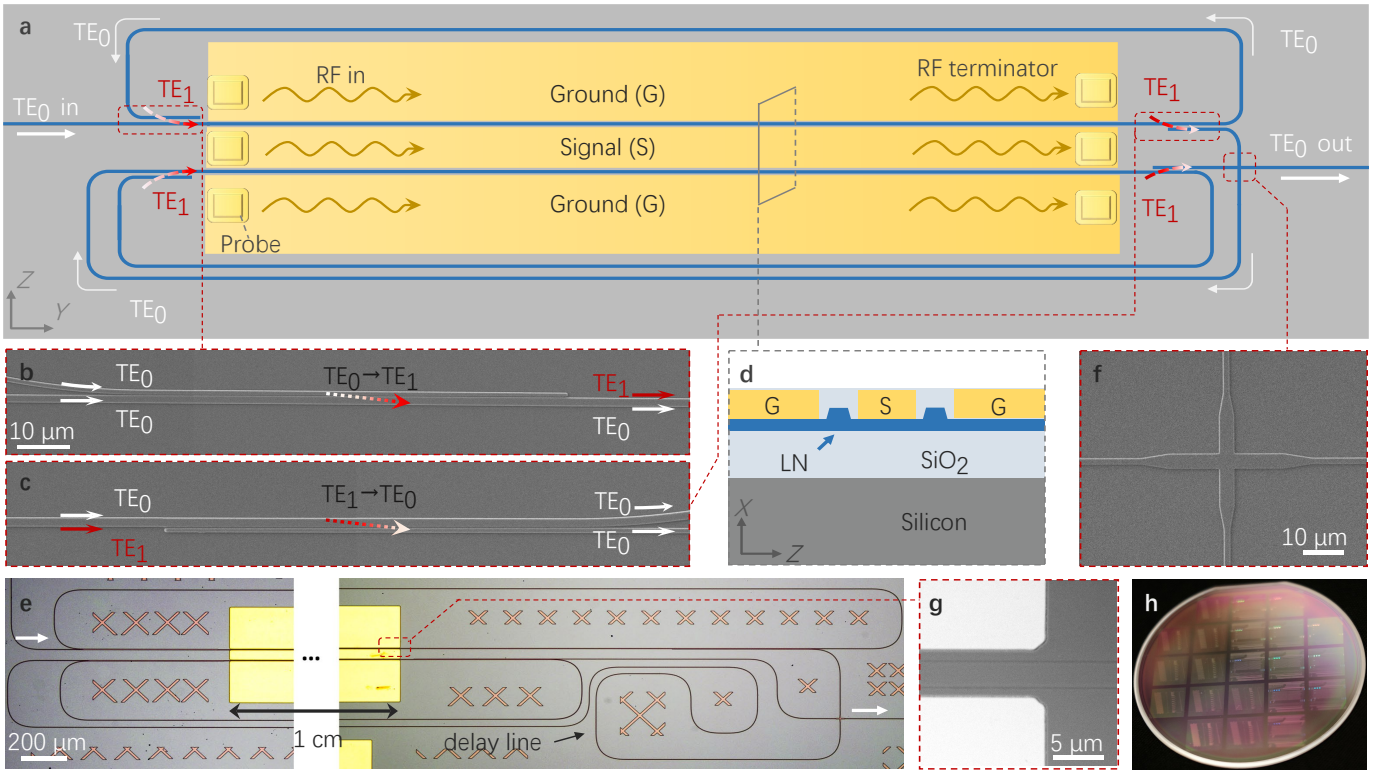

**Fig. 1 Schematic of the power-efficient on-chip EO comb generator.** (a) Device layout, where light travels through the phase modulation region for a total of 4 round trips via multiple TE$_0$/TE$_1$ mode conversions. (b-c) Scanning electron microscope (SEM) images of the adiabatic mode multiplexers. (d) Cross-sectional schematic of the modulation area. (e) Microscope image of the full EO comb generator, consisting of a 1-cm-long modulation region and engineered waveguide delay lines. (f-g) SEM images of the low-loss waveguide crossing (f) and the EO modulation area (g). (h) Camera image of a fabricated 4-inch LNOI wafer.

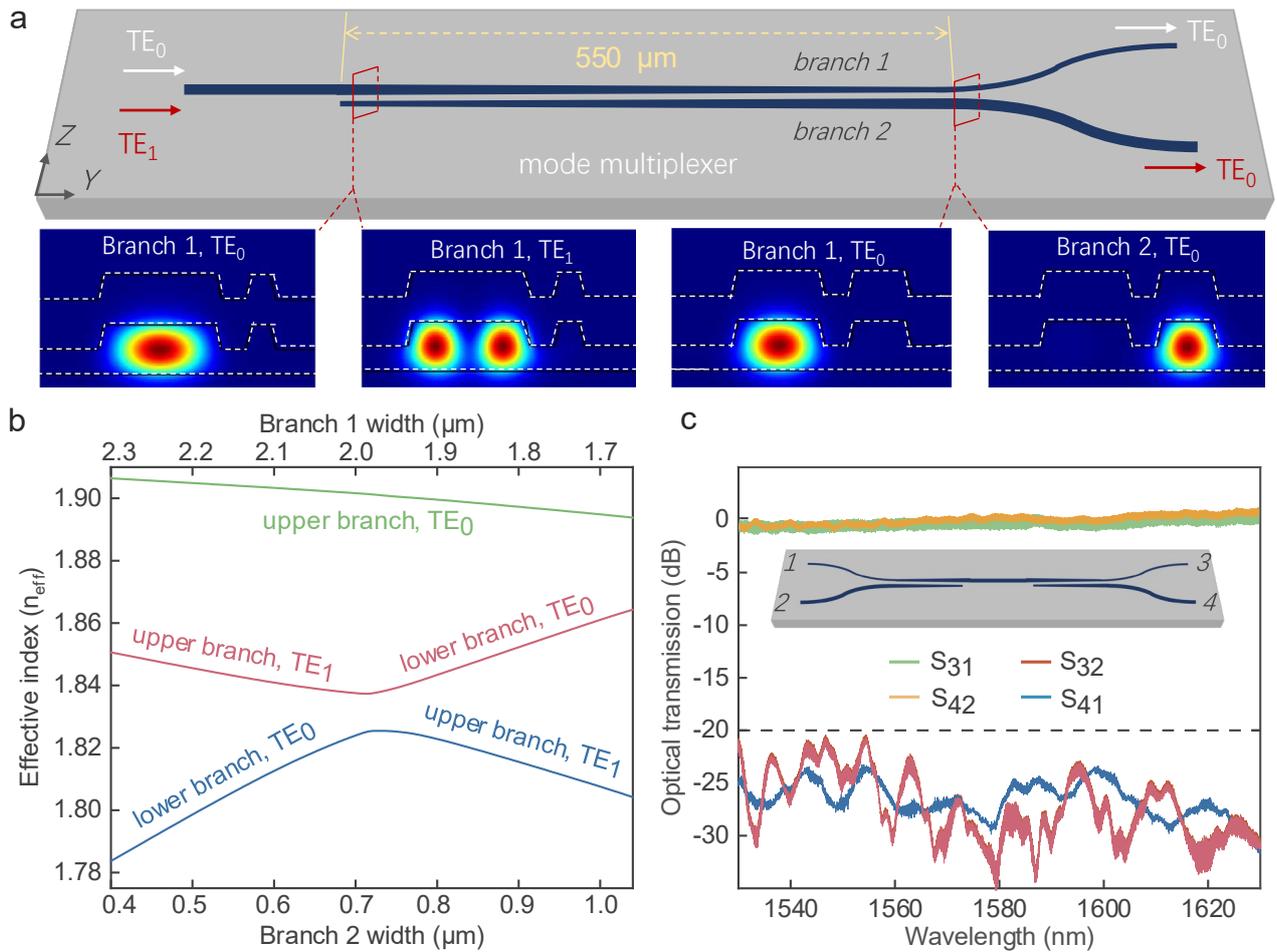

**Fig. 2 Design and characterization of the broadband adiabatic mode multiplexer.** (a) Device schematic, where input $TE_0$ and $TE_1$ modes are demultiplexed into $TE_0$ modes in the upper and lower output branches respectively. Insets show the corresponding simulated eigenmode profiles of the coupled-waveguide system at different locations. (b) Effective index ($n_{eff}$) evolution for the three lowest-order modes along our adiabatic mode multiplexer, at a wavelength of 1550 nm. (c) Measured optical transmission ($S_{31}$ and $S_{42}$) and crosstalk ($S_{32}$ and $S_{41}$) of a 2 × 2 cascaded multiplexer structure (inset), showing low optical losses of < 0.4 dB and high extinction ratios of > 20 dB from 1530 nm to 1630 nm.

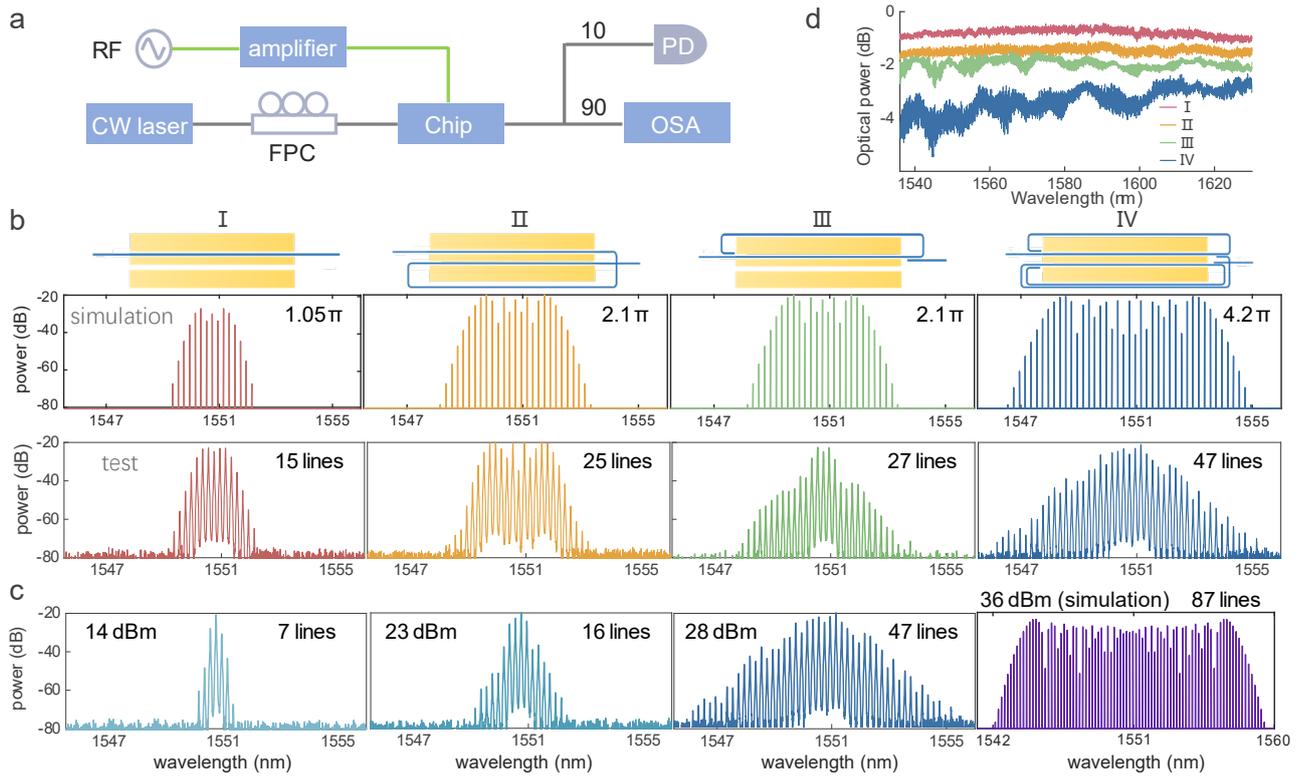

**Fig. 3 Power-efficient broadband on-chip EO comb generation.** (a) Experimental setup. (b) Optical frequency comb spectra generated from different types of EO comb generators at the same RF drive power. Top row: schematics of the four device types; middle row: simulated EO comb spectra; bottom row: measured EO comb spectra. (c) Comb spectra evolution at increasing RF driving power levels (14-28 dBm, measured; 36 dBm, simulated). (d) Optical transmission spectra of the four different EO comb generators.

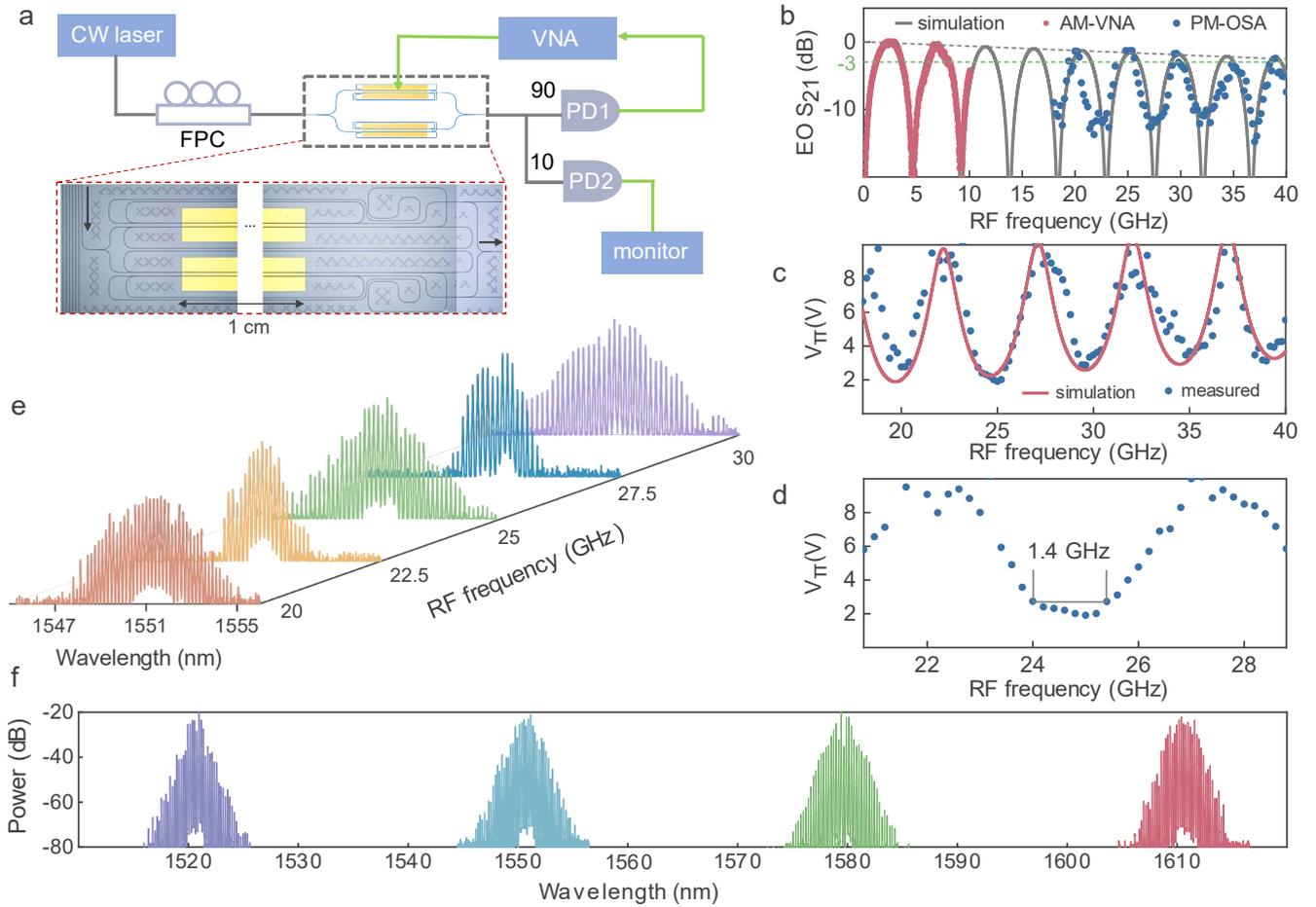

**Fig. 4 Tunability characterization of the EO comb generator.** (a) Experimental setup for direct EO response measurement using an MZI formed by two multi-loop phase modulators (inset: microscope image of the device). (b) Simulated and measured EO $S_{21}$ response. (c) Measured and simulated RF $V_\pi$ as functions of RF frequency. (d) Zoom-in view of the measured RF $V_\pi$ near 25 GHz, showing a 3-dB tuning bandwidth of ~ 1.4 GHz. (e) Measured EO comb spectra at different RF driving frequencies, showing periodically occurring maximal comb span near 20, 25 and 30 GHz. (f) EO comb generation at different optical pump wavelengths, showing a wide tuning range of operation wavelength.